\documentclass[10pt]{IEEEtran}
\usepackage[dvips]{graphicx}
\usepackage{amssymb}
\newcommand{\qed}{\fbox{}}
\newcommand{\ve}[1]{ \mbox{\boldmath$#1$} }
\newcommand{\defeq}{\stackrel{\triangle}{=}}
\newcommand{\limlog}{\lim_{n \rightarrow \infty} \frac 1 n \log}
\newtheorem{example}{Example}
\newtheorem{definition}{Definition}

\newtheorem{lemma}{Lemma}
\newtheorem{corollary}{Corollary}
\newtheorem{theorem}{Theorem}

\newcommand{\E}{{\rm E}}
\newcommand{\G}{{\cal G}}
\newcommand{\R}{{\cal R}}

\newcommand{\VAR}{{\rm VAR}}
\newcommand{\COV}{{\rm COV}}
\newcommand{\Gen}{{\cal G}}

\newcommand{\Bat}{Bhattacharya }

\begin{document}
\title{Asymptotic Concentration Behaviors of \\
Linear Combinations of Weight Distributions on \\
Random Linear Code Ensemble} 

\author{Tadashi Wadayama
\thanks{T. Wadayama is with 
  Department of Computer Science, 
  Nagoya Institute of Technology, Nagoya, 466-8555, Japan
  (e-mail: wadayama@nitech.ac.jp)
  } }
 \maketitle
\begin{abstract}
Asymptotic concentration behaviors of linear combinations
of weight distributions on the random linear code ensemble are presented. 
Many important  properties of a binary linear code can be expressed as the
form of a linear combination of weight distributions such as
number of codewords, undetected error probability and upper bound on 
the maximum likelihood error probability.
The key in this analysis is the 
covariance formula of weight distributions of the random linear code ensemble, which
reveals the second-order statistics of a linear function of the weight distributions.
Based on the covariance formula, several expressions of the {\em asymptotic concentration rate},
which indicate the speed of convergence to the average, are derived.
\end{abstract}

\section{Introduction}
For a binary random code ensemble or a binary random linear code ensemble, 
the asymptotic behaviors of the first moment (expectation) of some properties of
interest have been studied extensively. For example, the error exponent derived by Gallager \cite{Gal1} 
is a celebrated consequence of such a first-moment analysis.
Recent advances in second-moment analysis on low-density parity check matrix ensembles \cite{BB05}, \cite{VR05} have encouraged studies on the second-order behaviors (fluctuation from the average) of the macroscopic properties of an ensemble, which had previously attracted little attention.

In this paper, asymptotic concentration behaviors of linear combinations
of weight distributions on the random linear code ensemble are presented. 
Many important  properties of a binary linear code can be expressed as the
form of a linear combination of weight distributions such as
number of codewords, undetected error probability and upper bound on 
the maximum likelihood (ML) error probability.
The key in this analysis is the 
covariance formula of weight distributions of the random linear code ensemble, which
reveals the second-order statistics of a linear function of the weight distributions.
Based on the covariance formula, several expressions of the {\em asymptotic concentration rate},
which indicate the speed of convergence to the average, are derived.

\section{Preliminaries}

\subsection{Ensemble, expectation, and covariance}
Let $\G$ be a set of binary $m \times n$ matrices where
$m$ and $n$ are positive integers.
Suppose that probability $P(H)$ is assigned for each matrix $H$ in $\G$,
where $P(H)$ is a probability mass function defined on $\G$ such that
$
\sum_{H \in \G} P(H) = 1,
$ and 
$
\forall H \in \G, P(H) > 0.
$
The pair $\{\G, P(H)\}$ can be considered as an {\em ensemble of matrices}.
Although it is an abuse of notation, for simplicity, 
we will not distinguish $\{\G, P(H)\}$ from $\G$.

Let $f(\cdot)$ be a real-valued function defined on $\G$, which can be 
considered as a {\em random variable}.
The expectation of $f(\cdot)$ with respect to the ensemble $\G$ is defined by
$
\E_{\G} [f] \defeq \sum_{H \in \G} P(H) f(H).
$
The variance of $f(\cdot)$ is given by
$
\VAR_{\G} [f] \defeq \E_{\G} [ f(H)^2 ] - \E_{\G} [ f(H) ]^2.
$
In a similar way, the covariance between two real-valued functions $f(\cdot), g(\cdot)$
defined on $\G$ is given by
\begin{equation}
\COV_{\G} [f,g] \defeq \E_{\G} [ f g ] - \E_{\G} [ f ] \E_{\G} [ g].
\end{equation}

Let $\{g_1(\cdot), g_2(\cdot), \ldots, g_n(\cdot)\}$ be a set of 
real-valued functions defined on $\G$, and let
$f(\cdot)$ be a linear combination of $g_i(\cdot)$:
$
f(H) \defeq \sum_{i=1}^n \phi_i g_i(H)
$
for $H \in \G$, where $\phi_i (i \in [1,n])$ are  real values.
The notation $[a,b]$ denotes the set of consecutive integers from $a$ to $b$.
It is easy to show that the variance of $f(\cdot)$ is given by
\begin{equation} \label{varcov}
\VAR_{\G}[f ] =  \sum_{i=1}^n \sum_{j = 1}^n \phi_i \phi_j \COV_{\G}[g_i, g_j],
\end{equation}
e.g., see \cite{probandcomp} for details.
\subsection{Weight distribution}

The weight distributions $\{A_1(\cdot),\ldots, A_n(\cdot) \}$, 
which can be considered as a set of  real-valued functions defined on  $\G$, is defined by
\begin{equation}
A_w(H) \defeq \sum_{\ve x  \in Z^{(n,w)}}I[H \ve x = 0^m],\quad w \in [0,n],
\end{equation}
for any $H \in \G$, where $Z^{(n,w)}$ denotes the set of all binary $n$-tuples with weight $w$.
The function $I[\cdot]$ is the indicator function
such that $I[condition] = 1$ if $condition$ is true; otherwise, it gives 0.
In the present paper, symbol shown in bold, such as $\ve x$, denote column vectors.

Let $C(H)$ be the binary linear code defined based on $H$,
namely,
$
C(H) \defeq \{\ve x \in F_2^n: H  \ve x = 0^m\},
$
where $F_2$ denotes the binary Galois field.
Many properties of $C(H)$ of interest 
can be represented by a linear combination of the weight distributions $\{A_w(\cdot)\}_{w=1}^n$.
Let $F(\cdot)$ be such a property of $C(H)$, which is expressed as 
$
F(H) \defeq \sum_{w=1}^n \Phi_w A_w(H)
$
for any $H \in \G$, where $\Phi_w (w \in [0,n])$ are real values.

For example, the undetected error probability of $C(H)$ 
can be expressed as a linear combination of the weight distributions of $C(H)$
when it is used as an error detection code for a binary symmetric channel (BSC).
The expression is given by 
$
F(H) =  \sum_{w=1}^n A_w(H) \epsilon^w (1-\epsilon)^{n-w},
$
where $\epsilon$ is the crossover probability of the BSC.

In this setting, the property $F(\cdot)$ can be regarded as a 
random variable that takes a real value. It is natural to 
study its statistics such as expectation, variance for a given 
ensemble of binary matrices. 

\subsection{Random linear code ensemble}

In the present paper, we deal with an ensemble of binary matrices, which is called the {\em random linear code ensemble}.
\begin{definition}
The random linear code ensemble $\R_{n,m}$ contains all binary $m \times n$ matrices.
Equal probability 
$
P(H) = 1/2^{nm}
$
is assigned for each matrices in $\R_{n,m}$.
\hfill\qed
\end{definition}
Note that although the random linear code ensemble is actually an ensemble of matrices, it is regarded herein as an ensemble of binary linear codes.

The expectation of weight distributions of random ensemble is known \cite{Gal2} to be 
$
E_{\R_{n,m}}[A_w] = {n \choose w} 2^{-m} 
$
for $n \ge 1$.
The next theorem provides a closed formula of the covariance of weight distributions over the random linear code ensemble.
\begin{theorem}\label{covariancetheorem}
Assume a random ensemble $\R_{n,m}$.
The covariance of $A_{w_1}(\cdot)$ and $A_{w_2}(\cdot)$ is given by
\begin{eqnarray} \nonumber
&&\hspace{-1cm}\COV_{\R_{n,m}}[A_{w_1}, A_{w_2}]  \\
&=&\left\{
\begin{array}{ll}
0, & 0 < w_1, w_2 \le n, w_1 \ne w_2 \\
(1-2^{-m})2^{-m}{n \choose w}, & 0 < w_1 = w_2 \le n. \\
\end{array}
\right.
\end{eqnarray}
(Proof) The proof is given in Appendix.
\hfill\qed
\end{theorem}
The variance of the weight distributions of the random linear code ensemble 
has already been shown in \cite{modern}. Thus, the new contribution of this theorem is the case in which 
$\COV_{\R_{n,m}}(A_{w_1}, A_{w_2})  = 0$ when $w_1 \ne w_2$.
This theorem implies that the pair of random variables $A_{w_1}$ and $A_{w_2} (w_1 \ne w_2)$ 
is {\em pairwise independent}\footnote{Note that the set of random 
variables $\{A_{1},\ldots A_{n}\}$   are not mutually independent
because $\sum_{w=1}^n A_w(H) \ge 2^{n-m}-1$ holds for any instance $H$ in $\R_{n,m}$. }.

\section{Formulas on asymptotic concentration rate}

\subsection{Asymptotic behaviors of expectation}

\begin{definition}
Let $\G_n$ be an ensemble of binary $(1-R)n \times n$ matrices.
The parameter $R$, called the design rate, is a real value in the range of $0 < R < 1$. Suppose that $f(\cdot)$ is a real-valued function defined on $\G$. The {\em asymptotic exponent} of $\E_{\G}[f]$ is given by
\begin{equation}
\xi \defeq \lim_{n \rightarrow \infty} \frac 1n \log \E_{\G_n}[f]
\end{equation}
if the limit exists.
\hfill\qed
\end{definition}
Namely, asymptotically, $\E_{\G}[f]$ behaves like
$
E_{\G}[f(H)] \doteq 2^{\xi n}
$
where the notation $a_n \doteq b_n$ means that 
\[
\lim_{n \rightarrow \infty} (1/n) \log a_n = \lim_{n \rightarrow \infty} (1/n) \log b_n.
\]
In the present paper, a logarithm of base 2 is denoted by $\log$.

In the case of the random linear code ensemble,  it has been reported \cite{Gal2}  that 
\begin{equation}
\lim_{n \rightarrow \infty} \frac 1n \log \E_{\R_{n,(1-R)n}}[A_{\theta n}] 
= H(\theta) -(1-R),
\end{equation}
holds for $0 < \theta \le 1$, 
where $H(\cdot)$ is the binary entropy function defined by
$
H(x) \defeq -x \log x -(1-x) \log(1-x).
$
The parameter $\theta$ is called the {\em normalized weight}.

\subsection{Asymptotic concentration rate}

As the size of the matrix goes to infinity, the value of $f(\cdot)$ is often sharply concentrated around its expectation. The asymptotic concentration rate is defined as follows.

\begin{definition}
Let $\G_n$ be an ensemble of binary $(1-R)n \times n$ matrices, 
where $R$ is a real value in the range of $0 < R < 1$.
For a real-valued function $f(\cdot)$ defined on $\G_n$,
the asymptotic concentration rate (abbreviated as ACR) of $f(\cdot)$ is defined by
\begin{equation}
\eta \defeq \lim_{n \rightarrow \infty} \frac 1 n \log\left(\frac{\VAR_{\G_n}[f]}{\E_{{\G_n}}[f]^2} \right).
\end{equation}
if the limit exists.
\hfill\qed
\end{definition}

The following lemma explains the importance of the asymptotic concentration rate.
\begin{lemma}
Let $\eta$ be the asymptotic concentration rate of $f(\cdot)$.
For any positive real number $\alpha$,
\begin{equation}
\lim_{n \rightarrow \infty}\frac{1}{n}
\log Pr\left[\frac{f(H)}{\E_{\G_n}[f]} \notin (1-\alpha, 1+\alpha)   \right] \le \eta 
\end{equation}
holds if $\E_{\G_n}[f] > 0$ for any sufficiently large $n$. \\
(Proof) Based on the Chebyshev inequality,
the inequality 
\begin{equation} \label{chebychev}
Pr\left[ |f(H) - \E_{\G_n}[f]| >  c \sqrt{\VAR_{\G_n}[f]}  \right] \le \frac{1}{c^2}
\end{equation}
holds for any real number $c > 0$.
Suppose that $c$ is given by
\begin{equation}\label{cdef}
c = \frac{\alpha \E_{\G_n}[f] }{\sqrt{\VAR_{\G_n}[f]}}.
\end{equation}
where $\alpha$ is a positive real number. From the assumption $\E_{\G_n}[f] > 0$,
it is easy to verify that $c$ becomes positive.
Substituting (\ref{cdef}) into (\ref{chebychev}), we have
\begin{equation}
Pr\left[ |f(H) - \E_{\G_n}[f]| > \alpha \E_{G_n}[f]  \right] \le \frac{\VAR_{\G_n}[f]}{\alpha^2 E_{{\G_n}}[f]^2}.
\end{equation}
Due to the assumption $\E_{\G}[f(H)] > 0$, the above inequality can be rewritten in the following form:
\begin{equation}
Pr\left[ \frac{f(H)}{\E_{\G_n}[f]} \notin (1-\alpha, 1+\alpha)   \right] \le \frac{\VAR_{\G_n}[f]}{\alpha^2 \E_{{\G}}[f]^2}.
\end{equation}
Considering the asymptotic exponent of the above equation, we obtain the claim of the lemma.
\hfill\qed
\end{lemma}

From the asymptotic concentration rate, we can clarify
the probabilistic convergence behavior of $f(\cdot)$.
If $\eta<0$ holds, ${f(H)}/{E_{\G_n}[f]}$ converges to 1 in probability 
as $n$ goes to infinity. This means that $\eta<0$ is 
a sufficient condition of the convergence in probability.
The asymptotic concentration rate indicates the speed of this convergence

\begin{example}
The variance of the weight distributions of the random linear code ensemble 
is given by
\begin{equation}
\VAR_{\R_{n,(1-R)n}}[A_{\theta n}] = (1-2^{-(1-R)n}) 2^{-(1-R)n} {n \choose \theta n}.
\end{equation}
Therefore, the asymptotic exponent of the variance becomes
\begin{equation}
\limlog \VAR_{\R_{n, (1-R)n}}[A_{\theta n}] = H(\theta) - (1-R).
\end{equation}
From this exponent, we immediately have the asymptotic concentration rate of the weight distribution:
\begin{eqnarray} \nonumber
\eta&=& \limlog \frac{\VAR_{\R_{n,(1-R)n}}[A_{\theta n}]}{\E_{\R_{n,(1-R)n}}[A_{\theta n}]^2}  \\ \nonumber
&=& H(\theta) - (1-R) - 2\left(H(\theta) - (1-R)  \right) \\
&=& 1-R -H(\theta).
\end{eqnarray}
Let the minimum root of equation
$
1- R-H(\theta)  = 0
$
be $\theta_{GV}$, which is called the {\em relative Gilbert-Varshamov (GV) distance}.
Since $\eta<0$ holds in the range $\theta_{GV}< \theta < 1-\theta_{GV}$,
$A_{\theta n}(H)/E_{\R_{n,(1-R)n}}[A_{\theta n}] $ converges to 1 in probability as 
$n$ goes to infinity \cite{Barg}.
\hfill\qed
\end{example}

\subsection{ACR of a linear combination of weight distributions}

The goal of the present paper is to observe the asymptotic behavior of the variance of 
the linear combination defined in (\ref{fh}) of the weight distributions:
\begin{equation}\label{fh}
F(H) =  \sum_{w=1}^n \Phi_w A_w(H).
\end{equation}
The next theorem gives the asymptotic concentration rate of $F(H)$.
\begin{theorem}\label{acrformula}
Let $\G_n$ be an ensemble of binary $(1-R)n \times n$ matrices, which have the
following asymptotic first- and second-order behaviors:
\begin{eqnarray}
\E_{\G_n}[A_{\theta n}] &\doteq& 2^{n (H(\theta)+q(\theta))} , \\
\COV_{\G_n}[A_{\theta_1 n}, A_{\theta_2 n}] &\doteq& 2^{n \gamma(\theta_1,\theta_2)} .
\end{eqnarray}
The asymptotic concentration rate of $F(\cdot)$ defined in (\ref{fh}) is given by
\begin{eqnarray} \nonumber
\eta &=& \sup_{0 < \theta_1 \le 1} \sup_{0 < \theta_2 \le 1}
\left[\phi(\theta_1)+\phi(\theta_2) + \gamma(\theta_1, \theta_2)\right] \\
&-& 2\sup_{0 < \theta \le 1} \left[\phi(\theta)  +H(\theta) + q(\theta) \right],
\end{eqnarray}
where $\phi(\theta)$ is defined by
\begin{equation} \label{phidef}
\phi(\theta) \defeq \lim_{n \rightarrow \infty} \frac 1n \log \Phi_{\theta n}.
\end{equation}
(Proof) It is easy to verify that 
\begin{equation} \label{asyexpcet}
\lim_{n \rightarrow \infty} \frac{1}{n} \log \E_{\G_n}[F(H)] = \sup_{0 < \theta \le 1} 
\left[\phi(\theta)  +H(\theta) + q(\theta) \right] 
\end{equation}
holds. Using Eq.(\ref{varcov}),  we have
\begin{eqnarray} \nonumber
&& \hspace{-1.5cm} \lim_{n \rightarrow \infty} \frac 1 n \log \VAR_{\G_n}[F] \\ \label{asyvar}
&=& \sup_{0 < \theta_1 \le 1} \sup_{0 < \theta_2 \le 1}
\left[\phi(\theta_1)+\phi(\theta_2) + \gamma(\theta_1, \theta_2)\right].
\end{eqnarray}
Substituting (\ref{asyexpcet}) and  (\ref{asyvar}) into the definition of the ACR,
the theorem is proven. \hfill\qed
\end{theorem}

The next corollary is a special case of the above theorem for the random linear code ensemble.
\begin{corollary}
\label{acrrandom}
The ACR of $F(\cdot)$ defined in (\ref{fh}) over the random linear code ensemble $\R_{n,(1-R)n}$
is given by
\begin{equation}
\eta = \sup_{0 < \theta \le 1} \left[2 \phi(\theta)  +H(\theta)  \right]
-\sup_{0 < \theta \le 1} \left[2\phi(\theta)  +2H(\theta) \right] + 1-R,
\end{equation}
where $\phi$ is given in (\ref{phidef}). \\
(Proof) In the case of the random ensemble, $q(\theta)$ is given by
$
q(\theta)  =  - (1-R)
$
for $0 < \theta \le 1$.
From Theorem \ref{covariancetheorem}, we can derive 
the exponent of the covariance $\gamma(\theta_1, \theta_2)$, which is given by
\begin{equation}
\gamma(\theta_1, \theta_2) =
\left\{
\begin{array}{cc}
- \infty, & \theta_1 \ne \theta_2 \\
H(\theta) - (1-R), & \theta_1 = \theta_2, \\
\end{array}
\right.
\end{equation}
where $0 < \theta_1, \theta_2 \le 1$.
Plugging these functions into the formula in Theorem \ref{acrformula}, 
we obtain the claim of the corollary. \hfill\qed
\end{corollary}

\begin{example}
In this example, we will discuss the number of codewords in 
$C(H)$. Let us define
$
M(H) \defeq 1+\sum_{w = 1}^n A_w(H),
$
which is the number of codewords of $C(H)$.
In this case, we can see that $\Phi_w = 1$ holds for $1 \le w \le n$.
The asymptotic exponent of $M(H)$ is given by 
\begin{eqnarray} \nonumber 
\limlog \E_{\R_{n,m}}[M] &=& \sup_{0 < \theta \le 1}\left[ H(\theta) \right] - (1-R) \\
&=& R.
\end{eqnarray}
From the definition of $M(H)$, we immediately have
$
\phi(\theta) = 0,  0 < \theta \le 1.
$
Using Corollary \ref{acrrandom},  we obtain
\begin{eqnarray} \nonumber
\eta &=& \sup_{0 < \theta \le 1} \left[H(\theta) \right] -\sup_{0 < \theta \le 1} \left[2H(\theta)\right] + 1-R \\
&=& -R.
\end{eqnarray}
Since $R$ is a positive real number,  $M(H)/\E_{\R_{n,m}}[M]$ converges to 1 in probability 
for any $R >0$. \hfill\qed
\end{example}

In some cases, the asymptotic concentration rate can be written in 
a closed from without an optimization process required in Corollary \ref{acrrandom}.
\begin{theorem}
\label{acrformula2}
Assume the random linear code ensemble with design rate $R$.
Let $K_1, K_2$ be real positive constants that do not depend on $n$.
If $\Phi_w$ is expressed as
$
\Phi_w = K_1^{w} K_2^{n-w},
$
then the ACR of $F(H) =  \sum_{w=1}^n \Phi_w A_w(H)$ 
is given by
\begin{equation}
\eta = \log\frac{K_1^2+K_2^2}{(K_1+K_2)^2 }   +1-R.
\end{equation}
(Proof)  Using Theorem \ref{covariancetheorem} and the binomial theorem, we have
\begin{eqnarray} \nonumber
&&\hspace{-0.8cm}\VAR_{\R_{n,(1-R)n}}[F] \\ \nonumber
\hspace{-0.2cm}&=& \hspace{-0.3cm}\sum_{w_1=1}^n 
\sum_{w_2=1}^n (K_1^{w_1+w_2} K_2^{2n-w_1-w_2}) \COV_{\R_{n,(1-R)n}}[A_{w_1},A_{w_2}] \\ \nonumber
&=& \sum_{w=1}^n  (K_1^{2w} K_2^{2n-2w}) (1-2^{-m}) 2^{-m} {n \choose w} \\ \nonumber
&=& (1-2^{-m}) 2^{-m} \left(\sum_{w=0}^n  {n \choose w}(K_1^2)^w (K_2^2)^{n-w}    \right) \\ \nonumber
&-&(1-2^{-m}) 2^{-m} K_2^{2n} \\
&=& (1-2^{-m}) 2^{-m} \left(K_1^2+K_2^2   \right)^n-(1-2^{-m}) 2^{-m} K_2^{2n}.
\end{eqnarray}
Thus, the asymptotic exponent of $\VAR_{\R_{n,(1-R)n}}[F]$ is given by 
\begin{equation}
\limlog \VAR_{\R_{n,(1-R)n}}[F] = \log\left(K_1^2+K_2^2   \right)-(1-R).
\end{equation}
In a similar way, $\E_{\R_{n,(1-R)n}}[F]$ can be rewritten as follows:
\begin{eqnarray} \nonumber
\E_{\R_{n,(1-R)n}}[F] 
&=&
\sum_{w=1}^n (K_1^{w} K_2^{n-w}) \E_{\R_{n,(1-R)n}}[A_w]  \\ \nonumber
&=&
2^{-m} \left(\sum_{w=0}^n (K_1^{w} K_2^{n-w}) {n \choose w}  \right)  - 2^{-m} K_2^n \\ 
&=&
2^{-m} \left(K_1 + K_2 \right)^n  - 2^{-m} K_2^n.
\end{eqnarray}
This leads to the exponent of the expectation:
\begin{equation}
\limlog \E_{\R_{n,(1-R)n}}[F] = \log\left(K_1+K_2   \right)-(1-R).
\end{equation}
Substituting the above two equations into the definition of the ACR, 
we have the claim of the theorem.
\hfill\qed
\end{theorem}

\begin{example}
Assume the binary symmetric channel with crossover probability $\epsilon$.
The undetected error probability of $C(H)$ is given by
$
P_U(H) = \sum_{w=1}^n A_w(H) \epsilon^w \epsilon^{n-w}.
$
In this case,  the error exponent becomes
\begin{equation}
\lim_{n \rightarrow \infty} \frac{-1}{n} E_{\R_{n,(1-R)n}}[P_U] = 1-R.
\end{equation}
Since $\Phi_w = \epsilon^w \epsilon^{n-w}$ has the form stated in Theorem \ref{acrformula2}
(i.e., $K_1 = \epsilon, K_2 = 1-\epsilon$),
we can apply Theorem \ref{acrformula2}  and obtain 
$
\eta = \log(\epsilon^2 +(1-\epsilon)^2) + 1 - R.
$
This results suggests the existence of the convergence threshold $\epsilon^*$ for given $R$
such that $\epsilon^*$ separates the concentration regime and 
the non-concentration regime of $\epsilon$.
The root of $\log(\epsilon^2 +(1-\epsilon)^2) + 1 - R = 0$ becomes an upper bound of $\epsilon^*$.
Let $\epsilon'$ be the root of the equation
$
\log(\epsilon^{2} + (1-\epsilon)^{2})+1- R = 0.
$
Table \ref{estar} presents some values of $\epsilon'$ for  $0.1 \le R  \le 0.9$.
When $\epsilon > \epsilon'$, we have $ \log(\epsilon'^{2} + (1-\epsilon')^{2})+1- R <0$.
In such a region, $P_U(\cdot)$ concentrates around its average value in the limit as $n$ tends to infinity.
\begin{table}[htdp]
\caption{Roots of $\log(\epsilon^{2} + (1-\epsilon)^{2}) +1- R = 0$}
\begin{center}
\begin{tabular}{cc}
\hline
\hline
$R$ & $\epsilon'$ \\
\hline
0.1 &0.366047 \\
0.2 &0.307193 \\
0.3 & 0.259613 \\
0.4 &0.217375  \\
0.5 &0.178203 \\
0.6 & 0.140933 \\
0.7 &0.104872 \\
0.8 &0.069564 \\
0.9 &0.034687 \\
\hline
\end{tabular}
\end{center}
\label{estar}
\end{table}%

\hfill\qed
\end{example}

\section{ACR of the upper bound of ML error probability}

\subsection{\Bat bound}

In the following discussion, the binary symmetric channel 
with crossover probability $\epsilon$ is assumed for simplicity.
Assume that ML decoding is used in a decoder.
For a binary $m \times n$ parity check matrix $H$, the block error probability 
$P_e(H)$ can be upper bounded as follows:
\[
P_e(H)  \le \sum_{w = 1}^n A_w(H) D^w,
\]
where $D$ is called the \Bat parameter and is defined as
\[
D \defeq 2 \sqrt{\epsilon (1-\epsilon)}.
\]
The upper bound is called the {\em \Bat bound} \cite{Gal1} and has the form
of a linear combination of weight distributions.
Let us define
$
B(H) \defeq \sum_{w = 1}^n A_w(H) D^w.
$
It is expected that the statistics of $B(H)$ reflects the asymptotic  behavior 
of actual ML probability of an ensemble.

We first derive the asymptotic expression of the error exponent of
the \Bat bound in the case of the random linear code ensemble.
The expectation of $B(H)$ has the following closed form expression:
\begin{eqnarray} \nonumber
\E_{{\R}_{n,(1-R)n}} [B]  
&=& 
\sum_{w = 1}^n E_{{\R}_{n, (1-R)n}} [A_w(H)] D^w \\ \nonumber
&=&
\sum_{w = 1}^n {n \choose w} 2^{-(1-R)n} \left(2 \sqrt{\epsilon (1-\epsilon)} \right)^w \\ \nonumber
&=&
2^{-(1-R)n} (2 \sqrt{\epsilon (1-\epsilon)}+1)^n - 2^{-(1-R)n}.
\end{eqnarray}
Thus, the error exponent of $\E_{{\R}_{n,(1-R)n}} [B] $ is given by
\begin{eqnarray} \nonumber
&&\hspace{-2.5cm} \lim_{n \rightarrow \infty} \frac{-1}{n} \log \E_{{\R}_{n,(1-R)n}} [B]   \\ \label{errorexp}
&=& 1 -R -  \log\left(2 \sqrt{\epsilon (1-\epsilon)}+1 \right).  
\end{eqnarray}
This is a part of the error exponent function derived by Gallager \cite{Gal1} (see also \cite{Barg})
in the low-rate regime\footnote{It has been reported that this exponent is 
asymptotically tight if $R_x \le R \le R_{crit}$ \cite{Barg}.}.
Namely, the \Bat bound corresponds to the upper bound due to Gallager with the parameter 
$\rho = 1$ \cite{Gal1}.

In the following, we will examine the asymptotic concentration rate of the \Bat bound.
\begin{corollary}
The ACR of $B(H)$ is given by
\begin{equation}
\eta = \log\left(\frac{4 \epsilon (\epsilon-1) +1}{(2 \sqrt{\epsilon (1-\epsilon)}+1)^2} \right) + 1 - R.
\end{equation}
\end{corollary}
(Proof) By letting $K_1 = D$ and $K_2 = 1$ and using Theorem \ref{acrformula2}, we obtain
$
\eta = \log\left((D^2 +1)/(D+1)^2 \right) + 1 - R.
$
Substituting $D = 2 \sqrt{\epsilon (1-\epsilon)}$ into this equation, 
the corollary is proven.
\hfill\qed

\subsection{Expurgated bound}
We here consider the expurgated ensemble ${\R^*}_{n,(1-R)n}$, which 
can be obtained from ${\R}_{n,(1-R)n}$ 
by expurgating parity check matrices with $A_{\theta n}(H) \ne 0$ for $0 < \theta < \theta_{GV}, 1-\theta_{GV} < \theta \le 1$.
The asymptotic growth rate of the weight distributions is the same for the original and expurgated ensembles
when $\theta_{GV} \le \theta \le 1- \theta_{GV}$. However, $q(\theta)$ becomes $-\infty$ when $0 < \theta < \theta_{GV}, 1-\theta_{GV} < \theta \le 1$ in the case of the expurgated ensemble.

The error exponent of $\E_{{\R^*}_{n,(1-R)n}} [B] $ is given by
\begin{eqnarray} \nonumber
&&\hspace{-1.2cm} \lim_{n \rightarrow \infty} \frac{-1}{n} \log \E_{{\R^*}_{n,(1-R)n}} [B]   \\ \nonumber
&=& \min_{\theta_{GV} \le \theta \le 1-\theta_{GV}} \{1 - R - H(\theta) - \theta \log(2 \sqrt{\epsilon(1-\epsilon)}) \}.
\end{eqnarray}
If $\theta_{crit} \ge  \theta_{GV}$,
the minimum in the above equation is attained at $\theta = \theta_{crit}$, where
\begin{equation}
\theta_{crit}  \defeq \frac{2 \sqrt{\epsilon(1-\epsilon)}}{1+2 \sqrt{\epsilon(1-\epsilon)}}.
\end{equation}
In this case, the exponent coincides with the exponent given in Eq.(\ref{errorexp}).
Otherwise, $(\theta_{crit} <  \theta_{GV})$, the minimum occurs at $\theta = \theta_{GV}$. Therefore,  we have
\begin{equation}
\lim_{n \rightarrow \infty} \frac{-1}{n} \log \E_{{\R^*}_{n,(1-R)n}} [B]   \\ \nonumber
= - \theta_{GV} \log(2 \sqrt{\epsilon(1-\epsilon)}).
\end{equation}
if $\theta_{crit} < \theta_{GV}$. This exponent corresponds to the usual {\em expurgated exponent} for the BSC case (see also the discussion in \cite{Barg}).
The next corollary states the ACR of the upper bound of ML error probability in the case of $\theta_{crit} < \theta_{GV}$:
\begin{corollary}
If $\theta_{crit} < \theta_{GV}$, the ACR is given by $\eta = 0$.
(Proof) Since the expurgated ensemble can be obtained from
the original ensemble by removing a sub-exponential number of matrices, 
the exponent of the variance, i.e., $\gamma(\theta_1, \theta_2)$, 
takes  the same values  for the original and expurgated ensembles 
if $\theta_{GV} \le \theta_1, \theta_2 \le 1- \theta_{GV}$.
From Theorem \ref{acrformula}, we have 
\begin{eqnarray} \nonumber
\eta \hspace{-2mm}&=& \max_{\theta_{GV} \le \theta \le 1-\theta_{GV}} 
\left[H(\theta) + 2 \theta \log(2 \sqrt{\epsilon(1-\epsilon)})  \right] \\  \nonumber
&-& \hspace{-10mm}\max_{\theta_{GV} \le \theta \le 1-\theta_{GV}} 
\left[2H(\theta) + 2 \theta \log(2 \sqrt{\epsilon(1-\epsilon)})  \right]+1-R 
\end{eqnarray}
because $q(\theta) = -\infty$ for $\theta < \theta_{GV}$ in the case of the expurgated ensemble.
From the assumption $\theta_{crit} < \theta_{GV}$,   $2H(\theta) + 2 \theta \log(2 \sqrt{\epsilon(1-\epsilon)})$ 
is maximized at $\theta = \theta_{GV}$. Note that $-H(\theta_{GV}) + 1-R = 0$ holds.
Moreover, $H(\theta) + 2 \theta \log(2 \sqrt{\epsilon(1-\epsilon)})$ 
is also maximized at $\theta = \theta_{GV}$.
\hfill\qed
\end{corollary}
\section*{Appendix}
\subsubsection{Preparation of the proof of Theorem \ref{covariancetheorem}}
The second moment of the weight distribution for a given ensemble $\Gen$
is given by
\begin{eqnarray}
  \nonumber
&&\hspace{-1cm}\E_{\Gen}\left[A_{w_1}A_{w_2} \right]  \\ \nonumber
  &=&\hspace{-3mm}
  \E_{\Gen}\left[
   \sum_{\ve x \in Z^{(n,w_1)}} \sum_{\ve y \in Z^{(n,w_2)}}
    I[H \ve x = 0^m,H \ve y= 0^m] \right] \\ 
  &=& \hspace{-7mm}
   \sum_{\ve x \in Z^{(n,w_1)}} \sum_{\ve y \in Z^{(n,w_2)}}
   \E_{\Gen}\left[
    I[H \ve x = 0^m,H \ve y = 0^m] \right].
 \end{eqnarray}
For the case in which $\Gen = \R_{n,m}$, we obtain
\begin{eqnarray} \nonumber
&&\hspace{-1cm}E_{\R_{n,m}}[A_{w_1}A_{w_2} ]  \\ \label{eqcount}
  &=& \hspace{-5mm}\sum_{\ve x\in Z^{(n,w_1)}} \sum_{\ve y\in Z^{(n,w_2)}}
  \frac{\# \{H: H \ve x = 0^m, H \ve y = 0^m\}}{2^{mn}}.
\end{eqnarray}

Here, we encounter a problem of counting the matrices that satisfy 
both $H \ve x = \ve 0^m$ and  $H \ve y = \ve 0^m$.
Before solving this counting problem, we first introduce some notation.

Suppose that $w_1 > 0$ and $w_2 > 0$.
For a given pair $(\ve x, \ve y) \in Z^{(n,w_1)} \times Z^{(n,w_2)}$,
the index sets $I_1$, $I_2$, $I_3$, and $I_4$ are defined as follows:
$
I_1 \defeq \{k \in [1,n]: x_k = 1, y_k = 0 \},
I_2 \defeq \{k \in [1,n]: x_k = 1, y_k = 1 \},
I_3 \defeq \{k \in [1,n]: x_k = 0, y_k = 1 \},
I_4 \defeq \{k \in [1,n]: x_k = 0, y_k = 0 \},
$
where $\ve x=(x_1,x_2,\ldots,x_n)$ and $\ve y=(y_1,y_2,\ldots,y_n).$
The size of each index set is denoted by  $i _k = \# I_k (k = 1,2,3,4)$.
Let $\ve h=(h_1,h_2,\ldots,h_n)^t$ be a binary $n$-tuple (a row vector). 
The partial weight of $\ve h$ corresponding to an index set $I_k(k=1,2,3,4)$
is denoted by $w_k(\ve h)$, namely, 
$
  w_k(\ve h) \defeq \# \{j \in I_k: h_j = 1\}.
$

Since the index sets are mutually exclusive, the equation $i_1+i_2+i_3+i_4 = n$ holds 
and $i_2$ can take the integer values in the following range:
$
\max\{w_1+w_2-n,0  \} \le i_2\le \min\{w_1,w_2 \}.
$
The size of each index set can be expressed as 
$i_1 = w_1 - i_2$, $i_3 = w_2 - i_2$, 
$i_4= n - (w_1+w_2 - i_2)$.

The next lemma forms the basis for the proof of Theorem \ref{covariancetheorem}.
\begin{lemma} \label{l:fst}
For any $\ve x \in Z^{(n,w_1)}$ and $\ve y \in Z^{(n,w_2)} (0 < w_1, w_2 \le n)$,
the following equalities hold:
\begin{equation}
  \# \{\ve h \in F_2^n: \ve h \ve x = 0, \ve h \ve y = 0 \} = 
\left\{
  \begin{array}{ll}
  2^{n-2} & \ve x \ne \ve y, \\
    2^{n-1} & \ve x = \ve y.
  \end{array}
  \right.
\end{equation}
(Proof)
In the following, we prove the lemma
for the conditions $0 < w_1 \le w_2 \le n$. 
The proof for the opposite case $0 < w_2 \le w_1 \le n$ then 
follows immediately upon exchanging the variables $w_2$ and $w_1$ in the proof.

First, we will show that 
\begin{equation} \label{eq:fst}
\# \{\ve h \in F_2^n: \ve h \ve x = 0, \ve h \ve y = 0 \} =   2^{n-2} 
\end{equation}
 if $0 < w_1 \le w_2 \le n$ and $\ve x \ne \ve y$.
Let the support sets of $\ve x$ and $\ve y$ be $S(\ve x) \defeq \{i \in [1,n]: x_i = 1\}$ 
and $S(\ve y) \defeq \{i \in [1,n]: y_i = 1\}$, respectively.
The following three cases should be treated separately:
\begin{itemize}
\item Case (i): $0 < i_2 < w_1$ (i.e., $S(\ve x)$ and $S(\ve y)$  overlap but $S(\ve y)$ does not include $S(\ve x)$.)
\item Case (ii): $i_2 = 0$ (i.e.,  $S(\ve x)$ and  $S(\ve y)$ do not overlap.)
\item Case (iii): $i_2 = w_1$(i.e.,  $S(\ve y)$ includes $S(\ve x)$.)
\end{itemize}

First, we consider Case (i).
From the assumption that $0 < i_2 < w_1$,  it is clear that 
$I_1 \ne \emptyset$ (because $i_2 < w_1$), 
$I_2 \ne \emptyset$ (because $i_2 > 0$),
$I_3 \ne \emptyset$ (because $w_2  \ge w_1>i_2$).
For any $\ve h \in F_2^n$,
the equations $\ve h \ve x^t = 0$ and $ \ve h \ve y^t = 0$ hold if and only if
$
w_{i}(\ve h)\  \mbox{is even for } i = 1,2,3
$
or 
$
w_{i}(\ve h)\  \mbox{is odd for } i = 1,2,3.
$
Thus, the number of vectors satisfying the above condition is given by
\begin{equation}  \label{2n2}
N_{\ve h}  =
  2 \times 2^{i_1-1} \times 2^{i_2-1} \times 2^{i_3 -1}  \times 2^{i_4}  
= 2^{n-2},
\end{equation}
where $N_{\ve h}$ is defined by
$
N_{\ve h} \defeq  \# \{\ve h \in F_2^n: \ve h \ve x^t = 0, \ve h \ve y^t = 0 \}.
$
In the above derivation, we used the equalities: $w_1 = i_1 + i_2, w_2 = i_2 + i_3, i_4 = n-(w_1 + w_2 - i_2)$.
Note that Eq. (\ref{2n2}) (and Eqs. (\ref{eq:2nd}, )(\ref{eq:3rd}), and (\ref{eq:4th}) to be presented below) holds 
regardless of the size of $I_4$($i_4 = 0$ or $ i_4 >0$).

We now consider Case (ii). 
For this case, $I_1 \ne \emptyset$ (since $w_1>0$), $I_2 = \emptyset$ (since $i_2 = 0$) and 
$I_3 \ne \emptyset$ (since $w_2>0$).  
The equalities $\ve h \ve x = 0$ and $ \ve h \ve y= 0$ hold if and only if
$
w_{i}(\ve h)\  \mbox{is even for } i = 1,3
$
holds.
The number of vectors satisfying the condition is given by
\begin{equation}\label{eq:2nd}
N_{\ve h} 
= 2^{i_1 -1} \times 2^{i_3 - 1} \times 2^{i_4}  = 2^{n-2}.
\end{equation}

The final case is  Case (iii).
For this case, $I_1 = \emptyset$ (since $i_2 = w_1$), $I_2 \ne \emptyset$ (since $i_2 =w_1> 0$) and
$I_3 \ne \emptyset$ (since $\ve x \ne \ve y$ and $w_1 \le w_2$). 
These conditions lead to the condition:
$
w_{i}(\ve h)\  \mbox{is even for } i = 2,3
$
for $\ve h \ve x = 0, \ve h \ve y = 0$.
Again, $2^{n-2}$ $n$-tuples satisfy the above condition, namely,
\begin{equation} \label{eq:3rd}
N_{\ve h} 
= 2^{i_2 - 1} \times 2^{i_3- 1} \times 2^{i_4} 
= 2^{n-2}.
\end{equation}
Combining the above results for Cases (i), (ii), and (iii), we obtain Eq. (\ref{eq:fst}).

We then show that 
$
N_{\ve h}  =   2^{n-1} 
$
holds if $0 < w_1 = w_2 \le n$ and $\ve x = \ve y$.
For this case, we have $I_1 = \emptyset, I_2 \ne \emptyset, I_3 = \emptyset $(since $\ve x = \ve y$).
Thus, the equations $\ve h \ve x = 0, \ve h \ve y = 0$ hold if and only if
$
w_{2}(\ve h)\  \mbox{is even}.
$
The number of $n$-tuples satisfying the above condition is given by
\begin{equation} \label{eq:4th}
N_{\ve h}  
= 2^{i_2 - 1}  \times 2^{i_4} 
= 2^{n-1}.
\end{equation}
The proof of this lemma is completed.
\hfill\qed
\end{lemma}

\subsubsection{Proof of Theorem \ref{covariancetheorem}}
The proof of Theorem \ref{covariancetheorem} consists of two parts.
The first part corresponds to the case in which the covariance becomes zero.
The second part corresponds to the case in which the covariance becomes non-zero.

We commence with the first part of the proof.
Assume that $0 < w_1, w_2 \le n, \ve x \ne \ve y$.
From Lemma \ref{l:fst}, we obtain
\begin{eqnarray} \nonumber
&&\hspace{-2cm} \# \{H: H \ve x = 0^m, H \ve y = 0^m\} \\ \nonumber
&=& \prod_{k=1}^m \# \{\ve h \in F_2^n: \ve h \ve x = 0, \ve h \ve y = 0 \} \\ 
&=& 
2^{m(n-2)}.
\end{eqnarray}
Substituting into (\ref{eqcount}), we obtain 
\begin{eqnarray} \nonumber
\E_{\R_{n,m}}[A_{w_1}A_{w_2} ] 
  &=& \sum_{\ve x\in Z^{(n,w_1)}} \sum_{\ve y\in Z^{(n,w_2)}}
  \frac{ 2^{m(n-2)} }{2^{mn}} \\ \nonumber
 &=& 2^{-2m}{n \choose w_1} {n \choose w_2}  \\ 
 &=&  \E_{\R_{n,m}}[A_{w_1}] \E_{\R_{n,m}}[A_{w_2} ].
 \end{eqnarray}
The last equality is equivalent to $\COV_{\R_{n,m}}[A_{w_1}, A_{w_2}] = 0$.

We now consider the second part of the proof:
Assume that $\ve x = \ve y$.
From Lemma \ref{l:fst}, we have
$
\# \{H: H \ve x^t = \ve 0, H \ve y^t = \ve 0\} = 2^{m(n-1)},
$
and 
\begin{eqnarray} \nonumber
\E_{\R_{n,m}}[A_{w}^2]  \hspace{-2mm}
  &=& \hspace{-2mm} \sum_{\ve x\in Z^{(n,w)}} \sum_{\ve y\in Z^{(n,w)}}
  \frac{ I[\ve x = \ve y]2^{m(n-1)}   }{2^{mn}} \\ \nonumber
  &+& \hspace{-2mm}\sum_{\ve x\in Z^{(n,w)}} \sum_{\ve y\in Z^{(n,w)}}
  \frac{ I[\ve x \ne \ve y]2^{m(n-2)}   }{2^{mn}} \\ \nonumber
  &=&\hspace{-2mm} 2^{-m} {n \choose w} + 2^{-2m}\left({n \choose w}{n \choose w} - {n \choose w} \right)\\ 
        &=& \hspace{-2mm}\E_{\R_{n,m}}[A_{w}]^2+  2^{-m} {n \choose w} - 2^{-2m} {n \choose w} .
\end{eqnarray}
The last equality is equivalent to 
$
\COV_{\R_{n,m}}(A_{w}, A_{w}) = (1-2^{-m})2^{-m}{n \choose w}.
$
\hfill\qed

\section*{Acknowledgment}
The present study was supported in part by the Ministry of Education, Science, Sports, and Culture of Japan through a Grant-in-Aid for Scientific Research on Priority Areas (Deepening and Expansion of Statistical Informatics) No. 180790091.

\end{document}